\documentclass{agnspec}
\usepackage{graphics}
\usepackage{psfig,natbib}

\bibpunct{(}{)}{;}{a}{}{,}

\newcommand{\apj}{{ApJ}}
\newcommand{\apjs}{{ApJS}}
\newcommand{\apjl}{{ApJL}}

\newcommand{\aap}{{A\&A}}

\newcommand{\araa}{{ARA\&A}}
\newcommand{\mnras}{{MNRAS}}
\newcommand{\pasj}{{PASJ}}

\newcommand{\asca}{{\it ASCA}}
\newcommand{\sax}{{\it Beppo-SAX}}
\newcommand{\chandra}{{\it Chandra}}
\newcommand{\xmm}{{\it XMM-Newton}}
\newcommand{\hetg}{{\small HETG}}

\newcommand{\rgs}{{\small RGS}}
\newcommand{\syone}{Seyfert~1}
\newcommand{\sytwo}{Seyfert~2}
\newcommand{\nasa}{{\small NASA}}
\newcommand{\agn}{{\small AGN}}
\newcommand{\rrc}{{\small RRC}}
\newcommand{\uta}{{\small UTA}}
\newcommand{\ew}{\!EW}

\begin{document}
\title{Soft X-ray Spectra of \sytwo\ Galaxies}

\author{Masao Sako\inst{1,2}, Ali Kinkhabwala\inst{3}, Steven M. Kahn\inst{3},
        Ehud Behar\inst{3}, Frits Paerels\inst{3}, Ming Feng Gu\inst{2,4},
        Albert C. Brinkman\inst{5}, Jelle S. Kaastra\inst{5},
        \and Duane A. Liedahl\inst{6}}

\institute{
  Theoretical Astrophysics and Space Radiation Laboratory,
  Caltech, MC 130-33,
  Pasadena, CA 91125
\and
  Chandra Postdoctoral Fellow
\and
  Columbia Astrophysics Laboratory,
  550 West 120th Street, New York, NY 10027
\and
  Center for Space Research,
  Massachusetts Institute of Technology,
  Cambridge, MA 02139
\and
  SRON, the National Institute for Space Research,
  Sorbonnelaan 2, 3584 CA Utrecht, The Netherlands
\and
  Physics Department,
  Lawrence Livermore National Laboratory,
  P.O. Box 808, L-41, Livermore, CA 94550
}

\authorrunning{Sako et~al.}
\titlerunning{Soft X-ray Spectra of \sytwo\ Galaxies}

\maketitle

\begin{abstract}

  High resolution spectroscopic observations of Seyfert galaxies with
  \chandra\ and \xmm\ allows us to study the detailed ionization and thermal
  structures of the X-ray absorbing/emitting material in the circumnuclear
  environment.  The vast improvement in the spectral resolving power by more
  than an order of magnitude enables us, for the first time, to unambiguously
  distinguish the dominant line emission mechanisms and to measure its
  dynamical properties as well.  The X-ray band harbors spectral transitions
  from a wide range of ionization states, including valence-shell transitions
  in K-shell and L-shell ions from most cosmically abundant elements, as well
  as inner-shell transitions of iron and other mid-$Z$ elements, which can be
  probed through absorption measurements.  The X-ray spectrum, therefore,
  provides simultaneous velocity and column density constraints of highly
  ionized to only slightly ionized gas harbored in many of these systems.

  We summarize recent results that have emerged from observations of \sytwo\
  galaxies with the grating spectrometers onboard \chandra\ and \xmm.  We give
  particular emphasis to an empirical physical model that we have developed
  based on the observed spectra, and how it can be used for comparative
  studies with \syone\ galaxies to test the \agn\ unification scenarios.

\end{abstract}

\section{Introduction}

  In the past several years, X-ray observations have played an important role
  in the development of a ``unified model'' of \agn, in which the
  observational properties of the various classes (BL Lac objects, \syone\ and
  2, etc.) are explained solely in terms of their inclination angle with
  respect to the observer \citep{miller83, antonucci85, miller90,
  antonucci93}.  Such issues are naturally related to the understanding of the
  structure of matter immediately surrounding the central engine, where a
  variety of physical processes are expected to take place.  Spectral modeling
  and observations both suggest that the soft X-ray band should contain a
  wealth of information about the circumnuclear environment, which harbors
  regions ranging from relatively cool absorbing and reflecting media to hot
  and tenuous ionized regions.  X-ray spectra of many \agn\ exhibit strong
  emission lines, especially in \sytwo\ galaxies where the central continuum
  source is blocked by a torus of obscuring material and emission lines of
  large equivalent width are produced, both in the hard (2 -- 10 keV) and soft
  X-ray bands (0.3 -- 2 keV).  However, owing to the low spectral resolution
  spectra available prior to the deployment of \chandra\ and \xmm, physical
  parameters that may, in principle, be derived from the soft X-ray spectrum
  were not well-constrained.

  \asca\ and \sax\ observed strong soft X-ray line emission in many \sytwo\
  galaxies.  The nature of this line emission, however, has remained rather
  controversial, and models involving both photoionized and collisionally
  ionized plasmas yielded acceptable fits to the data (see, e.g.,
  \citealt{ueno94, iwasawa94, netzer97, turner97, griffiths98, sako00a}).

  \chandra\ and \xmm\ spectroscopic observations of Seyfert galaxies have
  provided us with a better understanding of the physical nature of the
  circumsource medium.  \citet{sako00b} have shown using \chandra\ \hetg\ data
  of Mrk~3 that the soft X-ray emission line spectrum is consistent with that
  produced in a warm absorbing medium seen in re-emission, providing further
  evidence that support the unified picture of \agn.  \citet{kinkhabwala02a,
  brinkman02a} and \citet{ogle02a} have performed quantitative analyses of the
  X-ray spectra of the archetypal \sytwo\ NGC~1068, and placed tight
  constraints on the column density and velocity distibution of the
  circumnuclear medium, as well as strict upper limits on the amount of
  collisionally ionized gas in this object.

%%%%%%%%%%%%%%%%%%%%%%%%%%%%%%%%%%%%%%%%%%%%%%%%%%%%%%%%%%%%%%%%%%%%%%%%%%%%%
\begin{figure}
  \centerline{\psfig{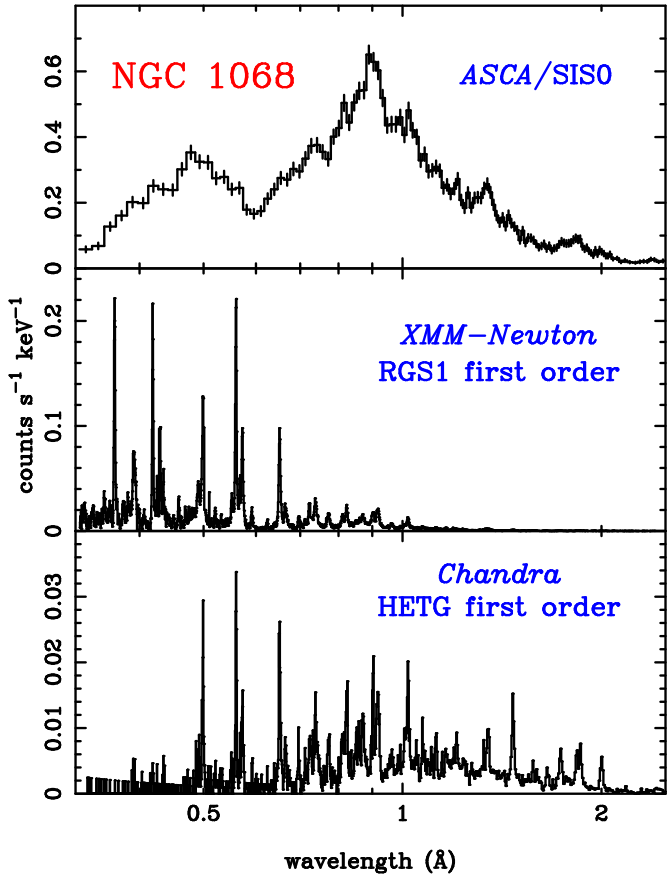}}
  \caption[]{The soft X-ray spectrum of NGC~1068 obtained with the \asca\ SIS0
  (top), the \xmm\ \rgs1 (middle), and the \chandra\ \hetg\ (bottom).  The
  spectrum is dominated by a forest of emission lines, which cannot be
  resolved with the resolving power capabilities of the \asca\ SIS ($E/\Delta
  E \sim 20$). \label{fig:1068}}
\end{figure}
%%%%%%%%%%%%%%%%%%%%%%%%%%%%%%%%%%%%%%%%%%%%%%%%%%%%%%%%%%%%%%%%%%%%%%%%%%%%%

  Although the X-ray spectra of \sytwo\ galaxies look very different from
  those of \syone, there is much overlap in the observable parameter space
  that characterize the physical nature of the circumsource medium.  In
  \syone, the central continuum source is used as a back-lighter to study the
  properties of the medium along the line of sight through absorption
  spectroscopy.  In \sytwo, on the other hand, where the direct view to the
  continuum source is blocked by the putative molecular torus, the properties
  of the absorbing medium can be studied by detailed investigation of light
  that is reprocessed and scattered into our line of sight.

  As mentioned above, absorption spectroscopy provides information pertaining
  to only a particular line of sight along which the observer happens to be
  looking.  This line of sight may or may not be representative of the entire
  circumnuclear region, and, as a consequence, it is not clear whether the
  derived parameters are representative of the global properties of the \agn.
  In principle, detection of emission lines superimposed on the background
  continuum provide some rough estimates of the covering fraction, since,
  unlike absorption lines, they are produced in regions that lie outside the
  line of sight as well, but a detailed investigation of the distribution of
  material is extremely difficult.  A similar complication exists in
  interpreting the spectra of \sytwo.  In these sources, the observed emission
  lines are produced in the entire circumsource medium, and the resulting
  spectrum is a sum over all the possible lines of sight from the central
  continuum source along the ionized medium.  The derived column densities
  are, therefore, biased towards regions of high covering fraction and column
  density (for reasons to be discussed in detail below).

  However, even given these complications, the information content of
  high-resolution X-ray spectra is certainly revolutionary.  In the remaining
  sections, we discuss what we can measure from the spectrum and the
  assumptions that go into the modeling and interpretation.  We identify and
  compare the parameter space spanned by spectra of both Seyfert 1s and 2s,
  and discuss how X-ray observations can be used to test the unified model of
  \agn.  Since much of the results on the X-ray spectral analyses of \syone\
  galaxies are discussed elsewhere in this {\it Proceedings}, here I will
  focus mainly on high resolution spectral data of \sytwo\ galaxies and its
  relation to those of \syone.

\section{Observations}

  Almost all of the X-ray bright, emission line dominated \sytwo\ galaxies
  have already been observed by the grating spectrometers on both \chandra\
  and \xmm.  These include the following sources: Mrk~3 \citep{sako00b},
  NGC~4151\footnote{Throughout this {\it article}, we simply refer to \syone\
  galaxies as objects in which the direct continuum radiation is visible in
  the soft X-ray region ($E \la 2 ~\rm{keV}$), and \sytwo\ as those in which
  the continuum below $E \sim 2 ~\rm{keV}$ is highly attenuated.  NGC~4151 was
  originally classified as a \syone.5 galaxy based on the optical emission
  line properties.  However, here we simply refer it as a \sytwo\ based on the
  X-ray continuum properties observed recently by \chandra.}  \citep{ogle00},
  the Circinus Galaxy \citep{sambruna01}, NGC~1068 \citep{kinkhabwala02a,
  brinkman02a, ogle02a}, and NGC~4507.  Since these objects were selected
  based on their soft X-ray brightnesses, their energy output is naturally
  dominated by \agn\ activity, and contamination from the host galaxy, e.g.,
  from starburst activity, is expected to be relatively small.  As shown
  below, the data also {\it require} that most, if not all, of the X-ray line
  emission is produced by \agn\ activity in all of the systems studied so far.
  In lower-luminosity \sytwo, the amount of starburst emission is expected to
  be more substantial.  High-resolution spectroscopic observations of such
  objects, however, still do not exist.

  In at least one object (NGC~1068), detailed imaging spectroscopy is possible
  with the \chandra\ grating spectrometers and the results are presented in
  \citet{brinkman02a} and \citet{ogle02b}, as well as in this {\it
  Proceedings} \citep{brinkman02b, ogle02b}.  In most of the other cases
  (Mrk~3, NGC~4151, and the Circinus Galaxy), some information regarding the
  spatial distribution is available as well.  High-resolution images show that
  the ionized media in \sytwo\ galaxies extend out to several hundred
  parsecs.  The minimum spatial scale that can be resolved by the \chandra\
  telescope corresponds to approximately 20 pc at the distance of the closest
  \sytwo\ in the sample (the Circinus Galaxy -- $1\arcsec \sim 20 ~\rm{pc}$ at
  $D = 4 ~\rm{Mpc}$).

\begin{table}[htbp]
  \begin{center}
    \caption[List of high-resolution spectroscopic observations of \sytwo\ galaxies.]
             {\label{tbl:sy2}
             List of High-resolution Spectroscopic Observations of \sytwo\ Galaxies.}
    \begin{tabular}[htbp]{llll}
      \\ \hline \hline
    Object name & RGS$^a$ & HETG$^a$ & LETG$^a$
      \\ \hline \hline
        NGC~1068 & $\surd$       & $\triangle$ & $\surd$ \\
        Mrk~3    & $\triangle$   & $\surd$     & $\times$    \\
        Circinus & $\triangle$   & $\surd$     & $\times$ \\
        NGC~4151 & $\triangle$   & $\surd$     & $\triangle$ \\
        NGC~4945 & $\triangle^b$ & $\times$    & $\times$ \\
        NGC~4507 & $\triangle$   & $\triangle$ & $\times$ \\
      \hline \hline
    \end{tabular}
  \end{center}
  \vspace{0.1in}
  \footnotesize
  $^a$ $\surd = $ the observation exists and the results are already
    published; $\triangle = $ the observation exists but the results are not
    yet in press at the time of writing (June 2002); $\times = $ observation
    does not exist at the time of writing \\
  $^b$ The EPIC data of a 24 ksec \xmm\ observation is presented in
    \citet{schurch02}, but the \rgs\ data have not yet been published.
\end{table}

\section{Spectroscopic Diagnostics and the Observables}

  In this section, we discuss some of the spectroscopic diagnostics useful for
  analyzing and interpreting high-resolution \syone\ and 2 spectra.

\subsection{\syone}

  Most of the spectroscopic features observed in \syone\ galaxies are
  absorption features that are produced through either photoexcitation
  (bound-bound) or photoionization (bound-free) of ions by the background
  continuum.  From an atomic physics point of view, absorption spectroscopy is
  relatively simple compared to emission spectroscopy since, in most practical
  cases, it only requires the knowledge of transition wavelengths, oscillator
  strengths, and radiative decay rates of resonance transitions, all of which
  are relatively easy to calculate.  Accurate laboratory wavelength
  measurements for the strongest lines are also available for most important
  ions.

  One of the unique characteristics of the X-ray band is that absorption lines
  of most H-like and He-like ions can be detected up to the series limit.  In
  an optically thin case, the absorption line equivalent width ($\ew$) ratios
  correspond to their oscillator strength ratios.  As the column density
  increases, the Ly$\alpha$ (or He$\alpha$; $n = 2 \rightarrow 1$) line
  saturate first, then the Ly$\beta$ (and He$\beta$; $n = 3 \rightarrow 1$)
  line, and so forth.  At a much higher column density, continuum absorption
  start to become important.  Therefore, by measuring the $\ew$ of multiple
  lines and using the curve-of-growth method, the X-ray spectrum is capable of
  probing an extremely wide range in column densities and turbulent velocities
  simultaneously.

  Velocity shifts can also be measured with high-resolution spectrometers.
  With the high energy transmission gratings onboard \chandra, for example,
  velocity shifts as low as $\sim 100 ~\rm{km~s}^{-1}$ can be measured for
  single isolated lines.  When multiple lines from a single ionic species are
  present, velocities can be constrained to higher accuracy.

  In practice, the thermal state of the irradiated medium is very difficult to
  discern from a \syone\ spectrum.  First of all, an absorption spectrum is
  completely insensitive to the dominant excitation/ionization mechanism that
  maintain the observed level of ionization.  Although it is true that
  photoionization/photoexcitation contribute to some level in populating the
  excited states of ions (because we see it), there may also be a large
  thermal heat source, which could dominate the total energy budget.  Another
  reason is that the spectral signatures of photoionization-dominated plasma
  (e.g., narrow radiative recombination continua -- \rrc\ -- and strong
  forbidden lines relative to resonance lines in He-like ions) are difficult
  to discern in the presence of a bright underlying continuum, especially when
  absorption lines are also superimposed.  In at least one source for which an
  extremely high signal-to-noise spectrum is available (NGC~3783 --
  \citealt{kaspi02}), however, there are clear detections of a number of \rrc\
  from different ions, thereby demonstrating the dominance of photoionization
  over collisional ionization.

\begin{figure}
  \centerline{\psfig{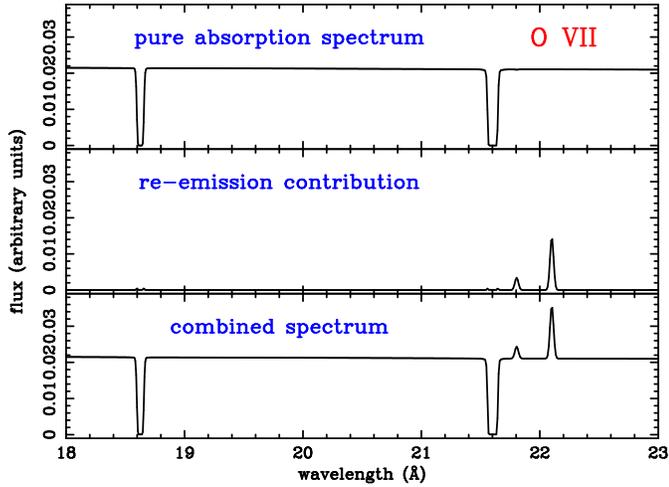}}
  \caption[]{Model \ion{O}{VII} spectra showing the effects of re-emission in
  an absorption spectrum.  This particular model assumes a covering fraction
  of $f = 0.1$ and a column density of $N_{\rm{\ion{O}{VII}}} = 10^{18}
  ~\rm{cm}^{-2}$.  The top panel shows a pure absorption spectrum with no
  re-emission.  The second panel shows the amount of re-emitted light that
  escapes the irradiated region in the absence of velocity gradients.  Note
  that the amount of resonance line radiation that escapes is much lower than
  in a pure recombining case (see Figure~\ref{fig:ovii}) due to
  self-absorption.  Note, however, that he results are sensitive to the
  density distribution of the medium.  \label{fig:ovii_reemit}}
\end{figure}

  The covering fraction can also be estimated by measuring the relative amount
  of absorption and emission observed in a given ion.  A model absorption
  spectrum of \ion{O}{VII} at a column density of $N_{\rm{\ion{O}{VII}}} =
  10^{18} ~\rm{cm}^{-2}$ is shown in the top panel of
  Figure~\ref{fig:ovii_reemit}.  The two absorption lines at $\lambda = 21.6$
  \AA\ and $\lambda = 18.6$ \AA\ are He$\alpha$ and $\beta$ transitions,
  respectively.  The middle panel shows the amount of re-emitted line emission
  that escapes a uniformly-filled ionization cone with a covering fraction of
  $f = 0.1$.  The forbidden line at $\lambda = 22.1$ \AA\ is produced solely
  through recombination cascades, and, in addition, almost always escapes the
  nebula because its oscillator strength is essentially zero.  Therefore, by
  measuring the absorption column density and the forbidden line intensity,
  one can estimate the total covering fraction of the absorbing medium.

\begin{table*}[htbp]
  \begin{center}
    \caption[Observable Parameters in \syone.]
             {\label{tbl:sy1par}
             Observable Parameters in \syone.}
    \begin{tabular}[htbp]{ll}
      \\ \hline \hline
    Parameter & Measurement Method
      \\ \hline \hline
      X-ray luminosity ($L_X$)    & -- direct measurement of continuum brightness \\
      ion column density ($N_i$)$^a$  & -- absorption line equivalent widths,
                                           and their
                                           ratios; curve-of-growth analysis \\
      turbulent velocity width ($v_{\rm turb}$)$^a$ &
                -- absorption line equivalent widths, and their ratios;
                   curve-of-growth analysis \\
             &  -- line profile analysis of individual lines or by
                   stacking (requires high signal) \\
      covering fraction ($f$)   & -- comparison of emission line intensities
                                     and absorption column densities \\
      line of sight velocity ($\Delta v$) & -- direct measurement of
                                               individual absorption line
                                               centroids or by stacking \\
      \hline \hline
    \end{tabular}
  \end{center}
  \vspace{0.1in}
  \footnotesize
  $^a$These two parameters are somewhat coupled.  See text for details.
\end{table*}

\begin{table*}[htbp]
  \begin{center}
    \caption[Observable Parameters in \sytwo.]
             {\label{tbl:sy2par}
             Observable Parameters in \sytwo.}
    \begin{tabular}[htbp]{ll}
      \\ \hline \hline
    Parameter & Measurement Method
      \\ \hline \hline
      X-ray luminosity ($L_X$)    & -- direct measurement is possible
                                      if the intervening medium is
                                       Thomson thin \\
      effective X-ray luminosity ($f \times L_X$)$^a$    &
          -- emission line intensities \\
      ion column density ($N_i$)$^a$  & -- relative line intensities;
                                           curve-of-growth analysis \\
      average turbulent velocity width ($\bar{v}_{\rm turb}$)$^a$ &
                -- relative line intensities; curve-of-growth analysis \\
      average bulk velocity field ($\Delta \bar{v}$) & -- direct measurement of
                                               individual absorption line
                                               centroids or by stacking \\
      size of ionized region ($R$) & -- direct measurement of spatial
                                        extent\\
      \hline \hline
    \end{tabular}
  \end{center}
  \vspace{0.1in}
  \footnotesize
  $^a$These parameters are somewhat coupled.  See text for details.
\end{table*}

%%%%%%%%%%%%%%%%%%%%%%%%%%%%%%%%%%%%%%%%%%%%%%%%%%%%%%%%%%%%%%%%%%%%%%%%%%%%%
%%%%%%%%%%%%%%%%%%%%%%%%%%%%%%%%%%%%%%%%%%%%%%%%%%%%%%%%%%%%%%%%%%%%%%%%%%%%%

  In addition to valence shell transitions of H-like and He-like ions and
  L-shell iron, the X-ray band is capable of probing material of lower charge
  states through inner-shell absorption spectroscopy.  For example, an
  unresolved transition array (\uta) of inner-shell $2p - 3d$ transitions of
  M-shell iron (\ion{Fe}{I} -- \ion{Fe}{XVI}) lie in the X-ray band between
  $\sim 15$ \AA\ and 17.5 \AA.  These features have been clearly detected in
  at least two objects -- an infrared quasar IRAS~13349+2348 \citep{sako01}
  and a \syone\ galaxy NGC~3783 \citep{kaspi01} -- as well as in the
  laboratory \citep{chenais00}.  Theoretical calculations of wavelengths and
  transition probabilities are presented in \citet{behar01}.

\subsection{\sytwo}

%%%%%%%%%%%%%%%%%%%%%%%%%%%%%%%%%%%%%%%%%%%%%%%%%%%%%%%%%%%%%%%%%%%%%%%%%%%%%
\begin{figure*}
  \resizebox{18cm}{!}{\rotatebox{90}{\includegraphics{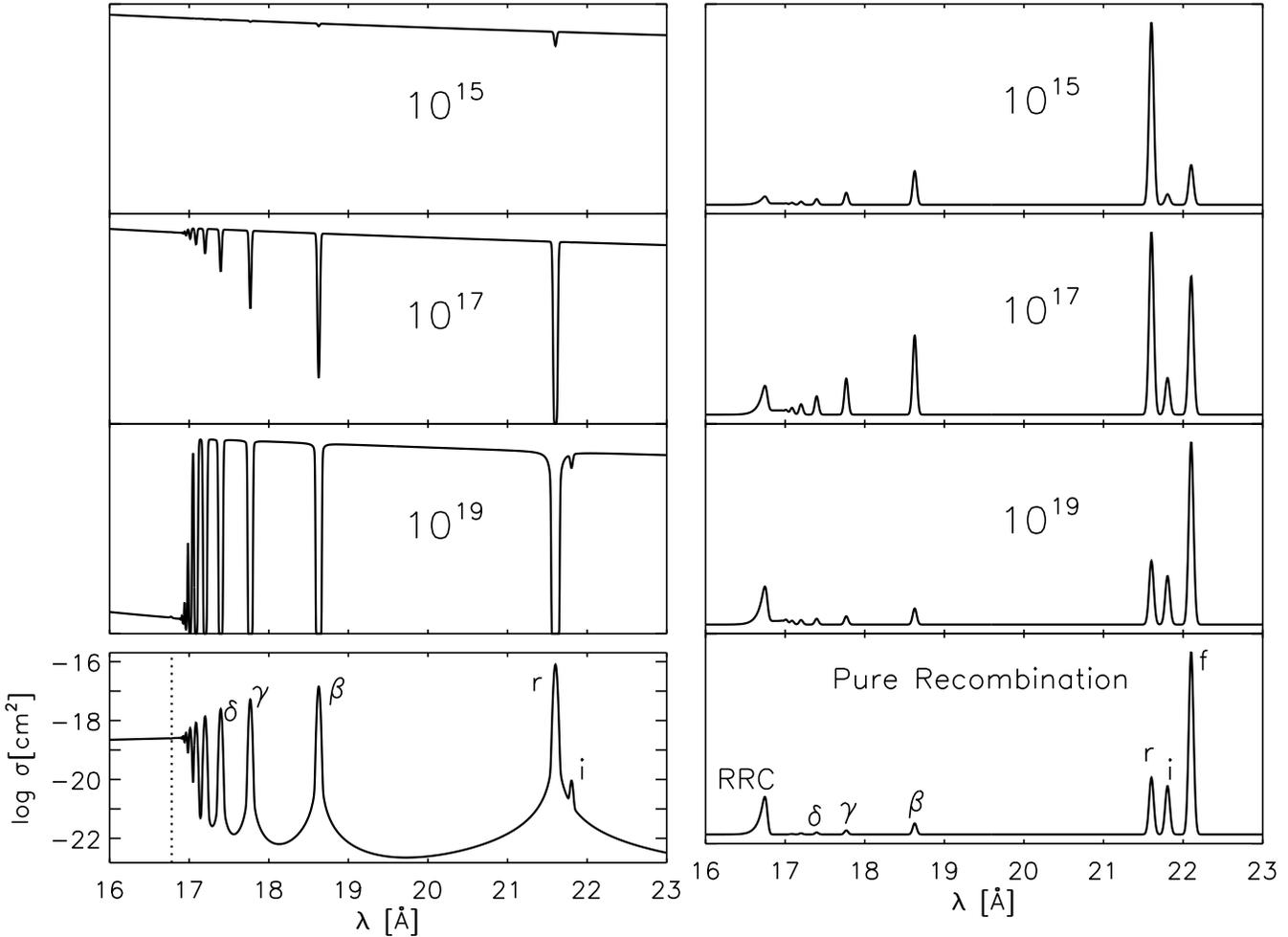}}}
  \caption[]{He-like \ion{O}{VII} absorption (left) and emission (right) model
  spectra at three different column densities.  For clarity, contribution from
  re-emission is not included in the absorption spectra.  At a column density
  of $N = 10^{15} ~\rm{cm}^{-2}$ (top), the resonance lines are optically
  thin.  In this case, roughly half of the emission line fluxes are produced
  through cascades following photoexcitation, while the other half are
  produced via recombination cascades (see text for details).  At a high
  column density ($N = 10^{19} ~\rm{cm}^{-2}$) (third panel), the resonance
  lines are completely saturated, while the edge is only mildly saturated.  In
  this case, recombination emission dominates over photoexcitation.  The
  resulting spectrum is, therefore, similar to a pure recombination spectrum
  shown on the lower-right panel, which is markedly different from the $N =
  10^{17} ~\rm{cm}^{-2}$ case.  The absorption cross section is shown on the
  lower-left panel.  The dotted vertical line designates the series limit.
  Model calculations are performed using a code available at {\tt
  http://xmm.astro.columbia.edu/photo/photo.html}.
  \label{fig:ovii}}
\end{figure*}
%%%%%%%%%%%%%%%%%%%%%%%%%%%%%%%%%%%%%%%%%%%%%%%%%%%%%%%%%%%%%%%%%%%%%%%%%%%%%

%%%%%%%%%%%%%%%%%%%%%%%%%%%%%%%%%%%%%%%%%%%%%%%%%%%%%%%%%%%%%%%%%%%%%%%%%%%%%
\begin{figure*}
  \resizebox{18.5cm}{!}{\rotatebox{90}{\includegraphics{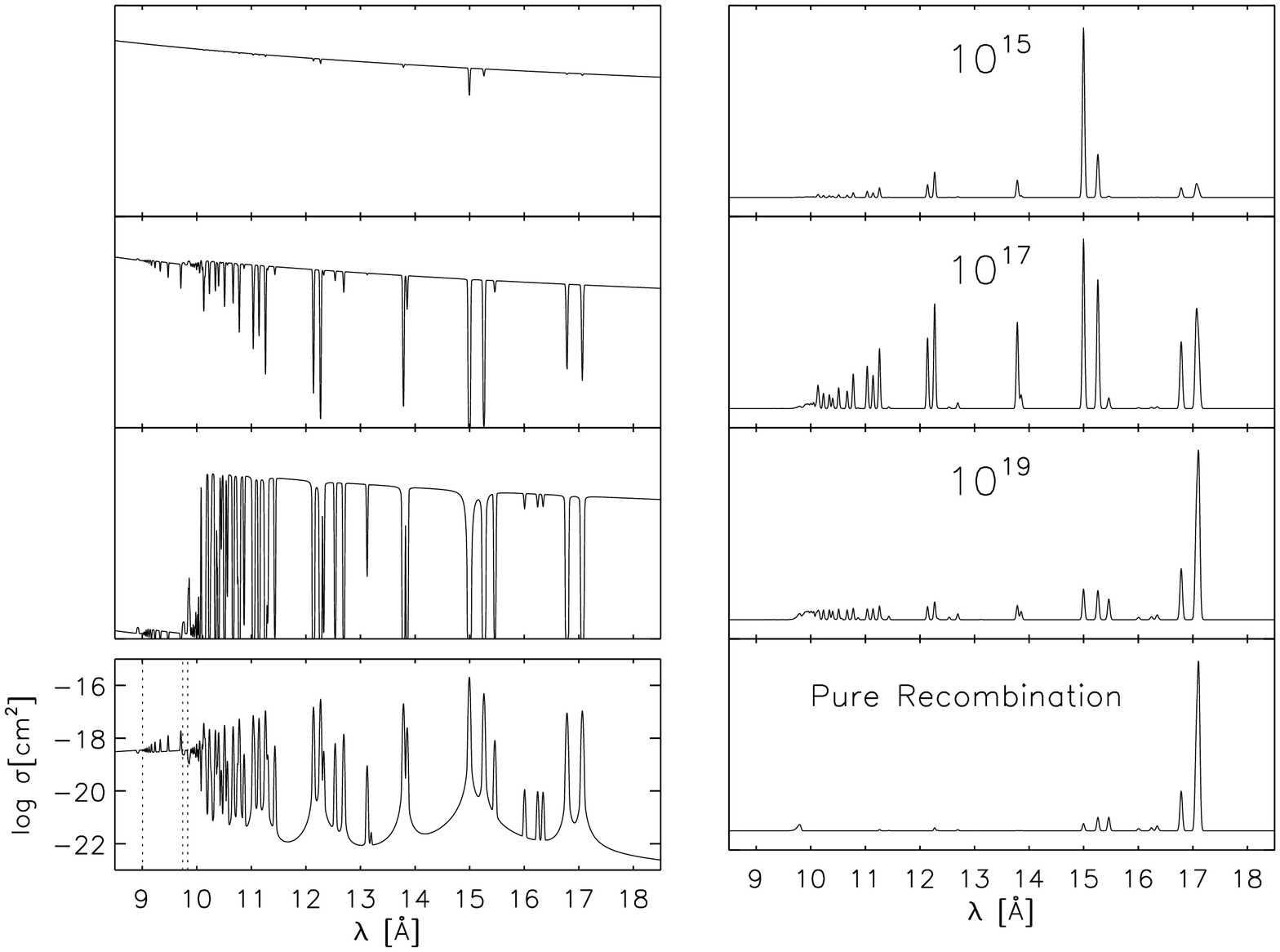}}}
  \caption[]{Same as in Figure~\ref{fig:ovii} for \ion{Fe}{XVII}.  The strong
  resonance lines at $\lambda = 15.01$ \AA\ and $\lambda = 15.24$ \AA\ are
  brightest at low column densities, while the lowest-level transition at
  $\lambda = 17.10$ \AA\ is stronger at high column densities. See text for
  details.  \label{fig:fexvii}}
\end{figure*}
%%%%%%%%%%%%%%%%%%%%%%%%%%%%%%%%%%%%%%%%%%%%%%%%%%%%%%%%%%%%%%%%%%%%%%%%%%%%%

  Most of the features in the spectra of \sytwo\ galaxies are produced through
  cascades following recombination and photoexcitation, which are inverse
  processes of those that produce absorption features in \syone\ spectra.

  As discussed in \citet{kinkhabwala02a}, careful measurements of emission
  line ratios provide simultaneous constraints on the column density and
  turbulent velocity, as in the case of \syone.  The method is analogous to
  the curve-of-growth method used in absorption spectroscopy, where the column
  density -- turbulent velocity degeneracy can be broken by measuring
  absorption equivalent widths of multiple lines from the same ionic species
  with different oscillator strengths and/or by measuring the depth of the
  photoelectric edge in cases where the column density is sufficiently high.

  When the direct view to the central continuum source is blocked by an
  obscuring medium, as in the case of \sytwo\ galaxies, these processes
  produce emission lines.  By computing photoionization and photoexcitation
  rates self-consistently, emission line intensities resulting from
  recombination cascades and cascades following photoexcitation can be
  calculated.  The predicted total emission line intensity ratios can then be
  compared with the data to yield column density and velocity width
  measurements.

%  Unlike in \syone, the ionized emission line medium lies perpendicular to our
%  line of sight, and distance estimates from the central source are also
%  possible for many of the systems.

  In H- and He-like ions, for example, approximately half of the total
  oscillator strength is in the lines and the other half in the continuum.
  This means that, in the optically thin limit, half of the total number of
  photons absorbed by the ion result in photoexcitations, while the other half
  result in photoionizations.  However, since the absorption cross section of
  strong resonance lines are several orders of magnitude larger than that of
  the continuum, they saturate at much lower column densities.  For example,
  the Ly$\alpha$ and He$\alpha$ lines saturate at a column of $N_{\rm ion}
  \sim 10^{15} ~\rm{cm}^{-2}$ and the continuum edge saturates at $N_{\rm ion}
  \sim 10^{19} ~\rm{cm}^{-2}$.  This is illustrated in Figure~\ref{fig:ovii},
  where we plot \ion{O}{VII} model spectra at three different column
  densities.  At a low column density (top panel), the intensity ratios of the
  resonance line series are roughly equal to their oscillator strength ratios.
  As the column density is increased, the lines saturate while the continuum
  is still optically thin.  Recombination line intensities increase roughly
  linearly with column density, while the photoexcitation intensities of the
  saturated lines increase only logarithmically (see also \citealt{behar02}).
  It is important to note that the He-like triplet line ratios at intermediate
  column densities look very similar to those in collisional ionization
  equilibrium, and need to be handled with caution.  Measurements of the
  higher order line intensities and the detection of the \rrc\ provide a more
  robust diagnostic for distinguishing the dominant emission mechanism
  \citep{kinkhabwala02a}.

  Similar diagnostics are possible with iron L ions (\ion{Fe}{XVII} --
  \ion{Fe}{XXIV}) as well.  In \ion{Fe}{XVII}, for example, the lowest lying
  level above the ground state produces the strongest recombination line at
  $17.10$ \AA \citep{liedahl90}.  This transition has an extremely low
  oscillator strength.  The transition at $15.01$ \AA\ is a strong resonance
  line and is efficiently excited via photoexcitation (or collisional
  excitation).  The $17.10$ and $15.01$ lines are analogous to the forbidden
  and resonance lines in He-like ions, respectively, and their line intensity
  ratio, therefore, provides constraints on the column density of
  \ion{Fe}{XVII}.  As in the He-like triplet, however, the $\lambda
  17.10/\lambda 15.01$ line ratio alone cannot be used to determine the
  emission mechanism unambiguously, but careful modeling/measurements of the
  weak lines provide robust discrimination.  The identification of an
  elaborate set of spectral diagnostics with iron L ions is in progress, and
  will be presented in \citet{gu02} and \citet{kinkhabwala02b}.

  Since the amount of line emission depends on the amount of reprocessed
  continuum radiation, the observed line intensities provide a measure of the
  X-ray continuum luminosity intercepted by the absorbing medium; i.e., the
  quantity $f \times L_X$, where $f$ is the covering fraction and $L_X$ is the
  X-ray luminosity.  In principle, careful measurements of line ratios can be
  used to infer the continuum shape as well, since photoexcitation line
  intensities are directly proportional to the local monochromatic continuum
  intensity.  This is, however, difficult in practice because the exact
  absorption column density is usually not known and radiative transfer
  effects may alter the model line ratios.

  Direct temperature estimates are also possible in \sytwo\ spectra by
  measuring the width of the \rrc.  The \rrc\ of \ion{C}{V} and \ion{Si}{XIV}
  are in isolated regions of the spectrum, while those of \ion{O}{VII} and
  \ion{O}{VIII} suffer from severe blending with several iron L lines.  Most
  of the other \rrc\ lie close to bright emission lines and may sometimes be
  difficult to measure the temperature with a spectrum of moderate statistical
  quality.  The temperature of the H-like and He-like CNO emission region in
  NGC~1068 is measured to be $kT \sim 30,000 - 45,000 ~\rm{K}$
  \citep{kinkhabwala02a}.

%%%%%%%%%%%%%%%%%%%%%%%%%%%%%%%%%%%%%%%%%%%%%%%%%%%%%%%%%%%%%%%%%%%%%%%%%%%%%
\begin{figure}
  \centerline{\psfig{figure=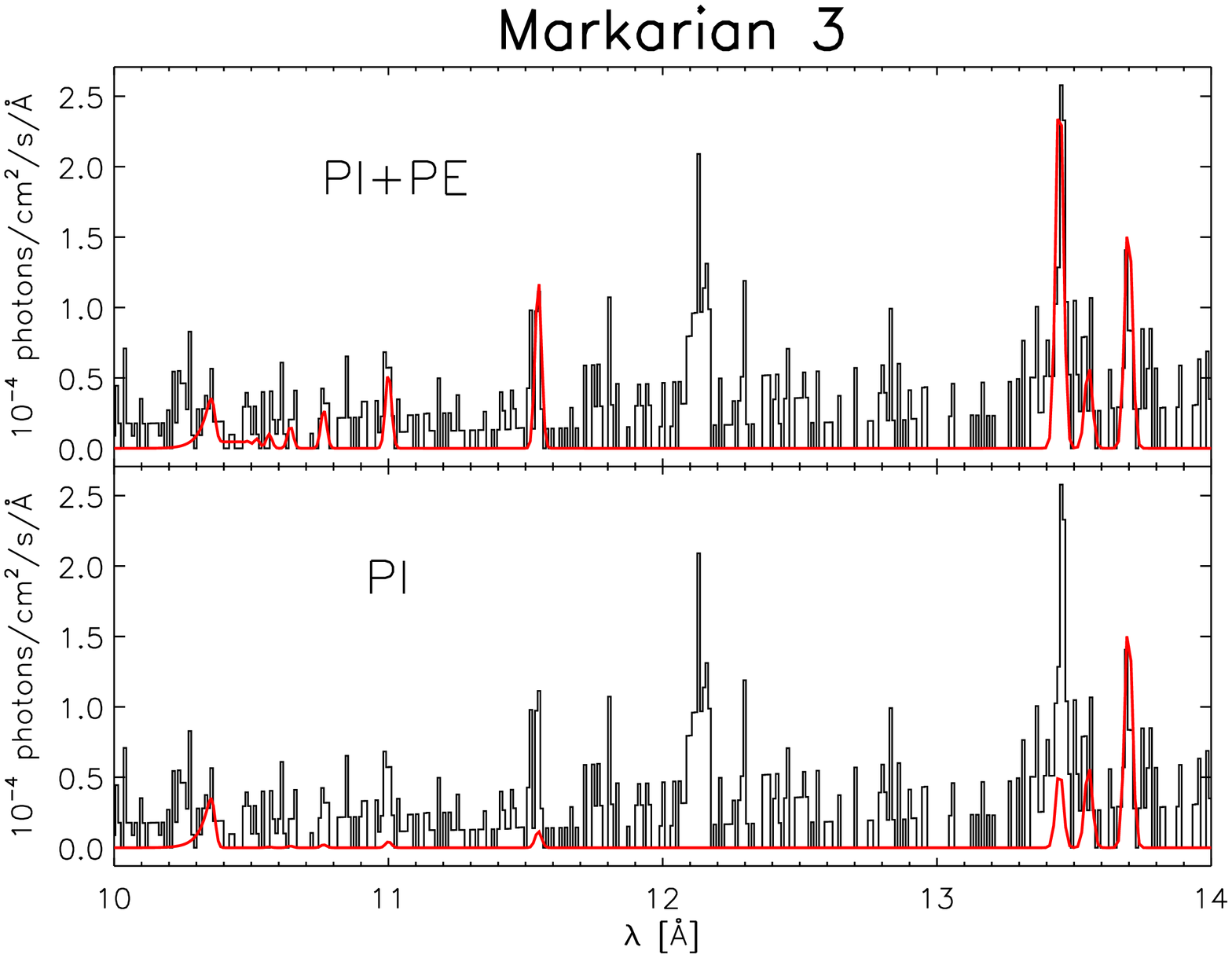,width=8.8cm,angle=90}}
  \vspace{0.2cm}
  \centerline{\psfig{figure=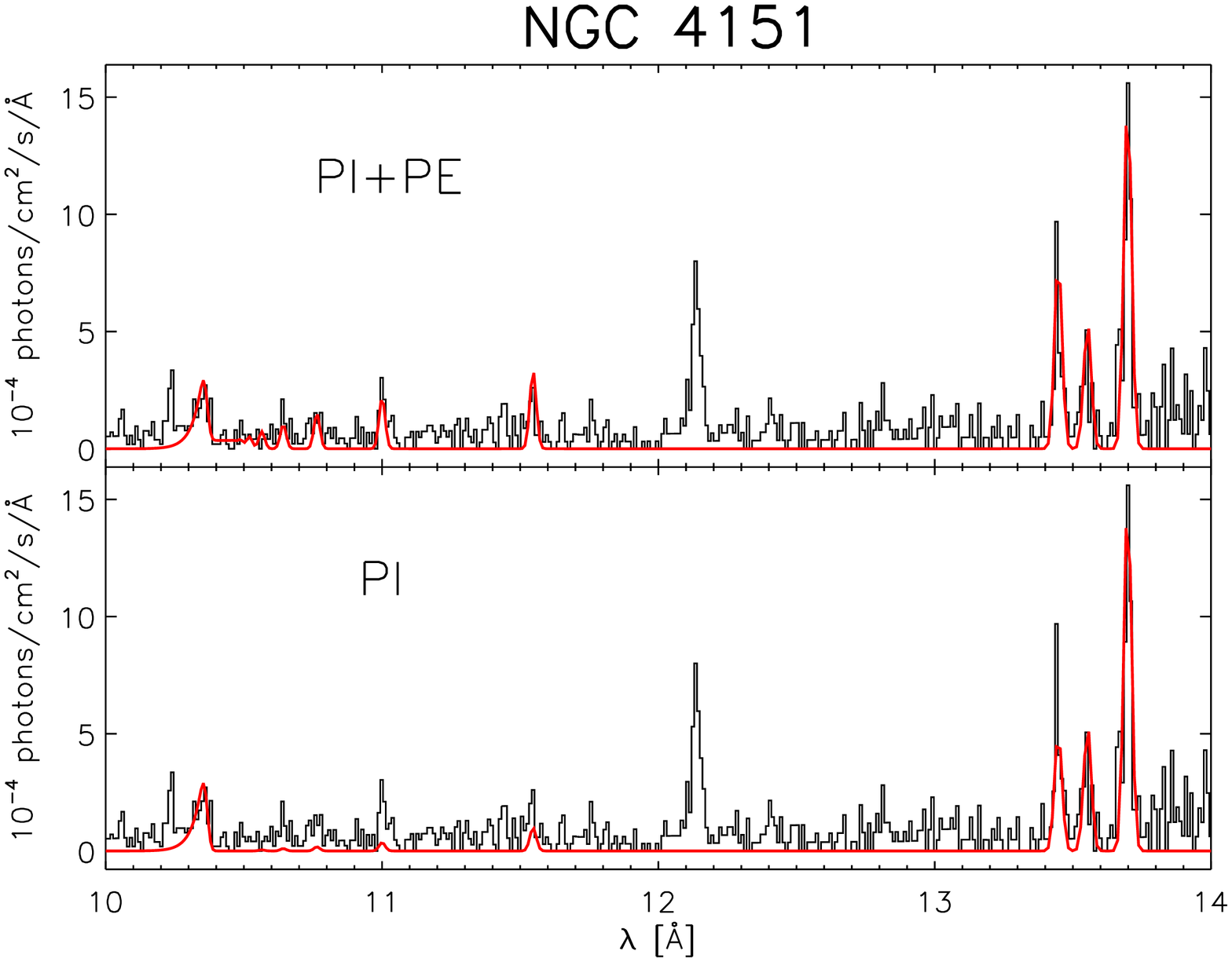,width=8.8cm,angle=90}}
  \caption[]{He-like \ion{Ne}{IX} spectra observed in Mrk~3 (top) NGC~4151
  (bottom).  The top panel of each figure shows the data with a \ion{Ne}{IX}
  model spectrum superimposed, which include emission from both
  photoionization and photoexcitation.  The bottom panel is the same, except
  the contribution from photoexcitation is excluded from the model.  The
  triplet lines are detected between 13 and 14 \AA.  Note the difference in
  the relative intensities of the forbidden and resonance lines.  In Mrk~3,
  the resonance line is brighter than the forbidden line, which suggests a
  rather low column density of Ne IX.  In contrast, the forbidden line in
  NGC~4151 is stonger. \label{fig:neix_1}}
\end{figure}
%%%%%%%%%%%%%%%%%%%%%%%%%%%%%%%%%%%%%%%%%%%%%%%%%%%%%%%%%%%%%%%%%%%%%%%%%%%%%

%%%%%%%%%%%%%%%%%%%%%%%%%%%%%%%%%%%%%%%%%%%%%%%%%%%%%%%%%%%%%%%%%%%%%%%%%%%%%
\begin{figure}
  \centerline{\psfig{figure=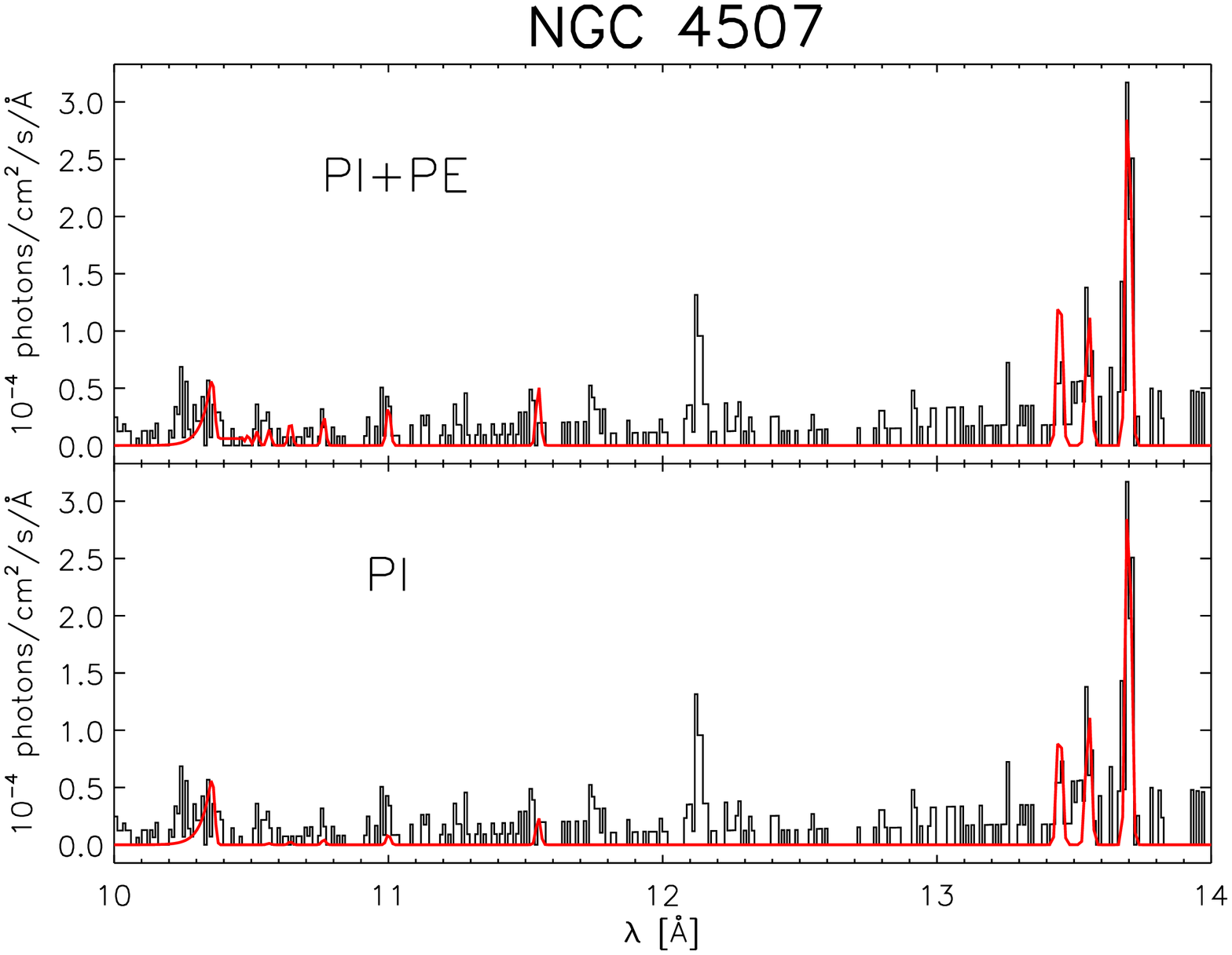,width=8.8cm,angle=90}}
  \caption[]{Same as in Figure~\ref{fig:neix_1} for NGC~4507.  Here, the
  observed forbidden line intensity is much stronger than that of the
  resonance line, which implies that photoionization dominates over
  photoexcitation (high column density). \label{fig:neix_2}}
\end{figure}
%%%%%%%%%%%%%%%%%%%%%%%%%%%%%%%%%%%%%%%%%%%%%%%%%%%%%%%%%%%%%%%%%%%%%%%%%%%%%

  Inner-shell absorption lines observed in \syone\ galaxies are extremely
  difficult to detect in re-emission, since the upper levels of many of these
  transitions are strongly autoionizing (i.e., low fluorescent yield).  A
  well-known exception is the ubiquitous iron K line complex near $6.4
  ~\rm{keV}$, which has an average fluorescent yield of $\sim 0.3 - 0.4$.
  Fluorescent line emission from other mid-$Z$ elements (Si, S, Ar, and Ca)
  has also been observed in several sources.  Inner-shell emission lines from
  low-$Z$ elements (C, N, O, and Ne) are probably too weak to be detected
  because of the $\sim Z^4$ dependence of the fluorescent yields.

  There is one important caveat in interpreting the measured column densities
  from a \sytwo\ spectrum.  As mentioned above, the emission line intensities
  are proportional to the quantity $f \times L_X$.  The line intensities are
  also monotonic functions of the column density, but the dependence is
  non-linear.  Therefore, in reality, the resulting spectrum is a complicated
  weighted sum over the entire ionized medium.  This implies that a measured
  ionic column density roughly corresponds to where the quantity $f \times
  N_i$ is a maximum; regions of small covering fraction and/or column density
  do not contribute significantly to the total X-ray line intensity.  This may
  not be a serious problem, since one can argue that those regions of
  extremely low column are of no particular interest.  It could, however, be
  detrimental for comparative studies with \syone, since absorption column
  densities are based on a single, unrepresantitive, line of of sight.

\section{Comparisons and Future Work}

  Although a detailed uniform spectral analysis of the \sytwo\ sample is in
  progress \citep{kinkhabwala02c}, rough comparisons of the derived physical
  parameters with \syone\ galaxies can already be made.

  NGC~3783 harbors, by far, the largest column density warm absorber observed
  to date.  From several \chandra\ \hetg\ observations, the total equivalent
  hydrogen column density in this object is estimated to be $N_H \sim 4 \times
  10^{22} ~\rm{cm}^{-2}$ \citep{kaspi02}.  Among the sources in the \sytwo\
  sample, NGC~4507 appears to have a warm absorber with comparable column
  density.  The observed \ion{Ne}{IX} triplet line ratios (see
  Figure~\ref{fig:neix_2}) imply an ionic column density of $N_{\rm
  \ion{Ne}{IX}} \sim 10^{18} ~\rm{cm}^{-2}$ or an equivalent hydrogen column
  density of $N_H \sim 2 \times 10^{22} ~\rm{cm}^{-2}$ assuming solar
  abundances.

  An example of an object with a small, but a detectable amount of absorption
  is Mrk~509, where we estimate the column density to be $N_H \sim 10^{21}
  ~\rm{cm}^{-2}$ \citep{pounds01}.  This is similar to the column density
  estimated by \citet{sako00b} for the \sytwo\ galaxy Mrk~3 (see, also
  Figure~\ref{fig:neix_1}).  A detailed investigation and comparisons of the
  sources listed in Table~\ref{tbl:sy2par} will be presented in
  \citet{kinkhabwala02c}.

  There are numerous other \syone\ objects with observed column densities that
  are much lower than $\sim 10^{21} ~\rm{cm}^{2}$.  These objects are much
  more difficult to detect from the \sytwo\ view, since, in these cases, the
  amount of reprocessed continuum radiation is small.  Deep exposures of X-ray
  faint \sytwo\ galaxies may allow us to place constraints on the column
  densities and to increase the parameter space in which detailed comparisons
  of \syone\ and 2 spectra are possible.

\begin{acknowledgements}

  {\small MS} and {\small MFG} were partially supported by \nasa\ through
  \chandra\ Postdoctoral Fellowship Award Number {\small PF}1-20016 and
  {\small PF}1-10014, respectively, issued by the \chandra\ X-ray Observatory
  Center, which is operated by the Smithsonian Astrophysical Observatory for
  and behalf of \nasa\ under contract {\small NAS}8-39073.  {\small AK}
  acknowledges support from an {\small NSF} Graduate Research Fellowship and
  \nasa\ {\small GSRP} fellowship.  The Columbia University team is supported
  by \nasa\ through the \xmm\ mission support and data analysis.  We
  gratefully acknowledge the permission by the Springer Verlag to use their
  A\&A \LaTeX{} document class macro.

\end{acknowledgements}

\end{document}